\documentclass[twocolumn,showpacs,amsmath,amssymb,prb,floatfix]{revtex4}

\usepackage{graphicx}
\usepackage{dcolumn}
\usepackage{bm}
\usepackage{epsfig}

\begin{document}

\title{Dynamical mean field theory for strongly correlated inhomogeneous
multilayered nanostructures}

\author{J.~K.~Freericks}
\affiliation{Department of Physics, Georgetown University,
             Washington, D.C. 20057}
\email{freericks@physics.georgetown.edu}
\homepage{http://www.physics.georgetown.edu/~jkf}

\date{\today}

\begin{abstract}
Dynamical mean field theory is employed to calculate the properties of
multilayered inhomogeneous devices composed of semi-infinite metallic lead 
layers coupled
via barrier planes that are made from a strongly correlated material (and can
be tuned through the metal-insulator Mott transition).  We find that the
Friedel oscillations in the metallic leads are immediately frozen in and don't 
change
as the thickness of the barrier increases from one to eighty planes.  We also
identify a generalization of the Thouless energy that describes the crossover
from tunneling to incoherent Ohmic transport in the insulating barrier.
We qualitatively compare the results of these self-consistent many-body 
calculations with the assumptions of non-self-consistent Landauer-based 
approaches to shed light on when such approaches are likely to yield good 
results for the transport.
\end{abstract}


\maketitle

\section{Introduction}

The fields of strongly correlated materials and of nanotechnology are
being united by work that investigates what happens when correlated materials 
are placed into inhomogeneous environments on the nanoscale.  This can
be accomplished by careful growth of strongly correlated materials with
molecular beam epitaxy or pulsed laser deposition, or it may be an
intrinsic property of some strongly correlated systems that display
either nanoscale phase separation, or nanoscale inhomogeneity. 
There are fundamental questions about these systems---what happens to the 
properties of the system when it has inhomogeneities on the nanoscale and how
does this spatial confinement modify the quantum-mechanical correlations?

We investigate a special case of a correlated nanostructure, where we
can carefully control the quantum confinement effects.  We take a semi-infinite
ballistic-metal lead and couple it to another semi-infinite ballistic-metal lead
via a strongly correlated barrier material (which is from one to eighty atomic
planes thick).  As the barrier is made thinner, the strongly correlated
system is being confined in one spatial direction between the metallic leads.
But the metallic leads induce a proximity effect on the barrier, which
can deconfine the correlated system.  Indeed, we will see that systems with a
single-plane barrier still display upper and lower Mott bands, but they also
have a low-energy low-weight peak to the density of states that
arises from the proximity-effect induced states that are localized
near the interfaces of the leads and the barrier.  As the barrier is made
thicker, this peak becomes a dip, which decreases exponentially with the
thickness.

We employ dynamical mean field theory (DMFT) in this work.  This allows us
to self-consistently calculate the properties of the inhomogeneous
system, including Friedel-like oscillations in the leads, and the
proximity-effect on the barrier.  We do not need to make any assumptions
about the kind of transport through this device, be it ballistic, diffusive, 
tunneling, or incoherent (via thermal excitations), since the DMFT
automatically incorporates all kinds of transport within its 
formalism\cite{potthoff_nolting_1999}.
We are, however, making one approximation in this approach---namely, we
make the assumption that the self energy remains local, even though
it can vary from plane to plane in the multilayered nanostructure.
Such an approximation should work fine for these inhomogeneous systems,
since the coordination number remains the same throughout the device
(and we are working in three dimensions).
This is to be contrasted with more conventional approaches to tunneling,
which assume a single-particle approach and employ a phenomenological
potential to describe the barrier region\cite{datta_1995}. The wavefunctions, 
transmission,
and reflection coefficients can be calculated, and then the transport 
analyzed, as in a Landauer-based approach.  In the DMFT calculations,
we determine the potential self-consistently (i.e., the self
energy) from the microscopic parameters of the Hamiltonian,
and the potential can vary with the energy of the scattering states. It is
not clear that a simple phenomenological potential can reproduce the
same kind of behavior via a conventional tunneling approach.

We assume each of the multilayer planes has translational invariance in 
the perpendicular $x$- and $y$-directions.  This allows us to use a 
mixed basis, Fourier transforming the two perpendicular directions
to $k_x$ and $k_y$, but keeping the $z$-direction in real space. Then
for each two-dimensional band energy, we have a quasi one dimensional
problem to solve, which has a tridiagonal representation in real
space, and can be solved with a renormalized perturbation 
expansion\cite{economou_1983}.
It is this mixed-basis representation (introduced by Potthoff and
Nolting\cite{potthoff_nolting_1999}) that allows us to solve this
problem.  By iterating our many-body equations, we can achieve a
self-consistent solution.

In addition to single-particle properties, we also evaluate $z$-axis
transport, perpendicular to the multilayers.  Thouless introduce the
idea of using the dwell time within the barrier to define a quantum
energy scale $\hbar/t_{dwell}$, which turned out to describe the
dynamics and transport of both ballistic metal and diffusive metal
barriers\cite{edwards_thouless_1972,thouless_1974}.  
The concept has been applied widely to the quasiclassical
theory of Josephson junctions as well\cite{dubos_etal_2001}.  
If we don't focus on the time
spent within the barrier, but instead try to extract an energy scale from
the resistance of a device, then we can generalize the Thouless energy
to the case of an insulating barrier, where the 
transport arises from
either tunneling or incoherent (thermally activated) processes. We
find that when this energy scale is on the order of the temperature, then
we have a crossover from tunneling to incoherent transport.
A short communication of this work has already appeared\cite{freericks_2004}.

The organization of this paper is as follows: in Section II, we present
a detailed derivation of the formalism and the numerical algorithms
used to calculate properties of nanostructures.  In Section III, we describe
the single-particle properties, focusing on the density of states and the
self energy.  In Section IV, we generalize the concept of the Thouless
energy, which is applied to charge transport in Section V.  We end
with our conclusions in Section VI.

\section{Formalism and numerical algorithms}

The Hamiltonian we consider involves a hopping term for the
electrons and an interaction term for the sites within the barrier
region (interactions can be added in the metal if desired to convert the 
leads from a ballistic metal to a diffusive metal, but we do not 
do so here).  For the interaction, we employ the Falicov-Kimball 
model\cite{falicov_kimball_1969}
which involves an interaction between the conduction electrons with
localized particles (thought of as $f$-electrons or charged ions) when the
conduction electron hops onto a site occupied by the localized particle.
We consider spinless electrons here, but spin can be included trivially
by introducing a factor of 2 into some of the results.
The Hamiltonian is
\begin{equation}
\mathcal{H}=-\sum_{ij}t_{ij}c^\dagger_ic_j+\sum_iU_i\left ( 
c^\dagger_ic_i- \frac{1}{2}\right )\left ( w_i-\frac{1}{2}\right )
\label{eq: hamiltonian}
\end{equation}
where $t_{ij}$ is a Hermitian hopping matrix, $U_i$ is the Falicov-Kimball
interaction, and $w_i$ is a classical variable that equals one if
there is a localized particle at site $i$ and zero if there is no
localized particle at site $i$ (a chemical potential $\mu$ is employed
to adjust the conduction-electron concentration).  Since we are considering 
multilayered
heterostructures, we assume that the hopping matrix is translationally invariant
within each plane, as well as the Falicov-Kimball interaction.  We let
the $z$-direction denote the direction where the system is allowed to have
inhomogeneity.  Then our translational invariance in the parameters
requires that $U_i=U_j$ if ${\bf R}_i-{\bf R}_j$ has a vanishing $z$-component.
Similarly, $t_{ij}=t_{i^\prime j^\prime}$ if ${\bf R}_i-{\bf R}_{i^\prime}$
and ${\bf R}_j-{\bf R}_{j^\prime}$ both have a vanishing $z$-component,
and ${\bf R}_i-{\bf R}_{j}={\bf R}_{i^\prime}-{\bf R}_{j^\prime}$.
But this requirement is quite modest, and allows for many complex
situations to be considered.  

We denote the planes with a given $z$-component
by a Greek label ($\alpha$, $\beta$, $\gamma$, ...).  Then our requirement
on the interaction is that $U_{\alpha}$ has a definite value for each
plane $\alpha$.  The hopping matrix can have one value $t_{\alpha}^\parallel$
for the hopping within the plane, and different values $t_{\alpha,\alpha+1}$
and $t_{\alpha-1,\alpha}$ for hopping to the plane to the right and
for hopping to the plane to the left, respectively.  For simplicity, we will
only consider nearest-neighbor hopping here, and we assume the lattice positions
${\bf R}_i$ all lie on the points of a simple cubic lattice (but we do not
have full cubic symmetry).

Because of the translational invariance within each plane, we can perform 
a Fourier transform in the $x$- and $y$-coordinates to the mixed basis
${\bf k}_x$, ${\bf k}_y$, and $\alpha$ (the $z$-component in real space).
We define the two-dimensional band structure, for each plane $\alpha$,
by 
\begin{equation}
\epsilon_\alpha^{2d}({\bf k}_x,{\bf k_y})=-2t_\alpha^\parallel
[\cos{\bf k}_x+\cos{\bf k}_y].
\label{eq: 2dbandstructure}
\end{equation}
The Green's function, in real space, is defined by
\begin{equation}
G_{ij}(\tau)=-\langle {\mathcal T}_\tau c_i(\tau)c^\dagger_j(0)\rangle,
\label{eq: g_def}
\end{equation}
for imaginary time $\tau$. The notation $\langle {\mathcal O}\rangle$ denotes
the trace ${\rm Tr} \exp(-\beta [\mathcal{H}-\mu{\mathcal N}]){\mathcal O}$
divided by the partition function ${\mathcal Z}={\rm Tr} \exp(-\beta 
[\mathcal{H}-\mu{\mathcal N}])$ and the operators are expressed in the 
Heisenberg representation ${\mathcal O}(\tau)=\exp(\tau [\mathcal{H}-\mu
{\mathcal N}]){\mathcal O}\exp(-\tau [\mathcal{H}-\mu{\mathcal N}])$. The
symbol ${\mathcal T}_\tau$ denotes time ordering of operators, with
earlier $\tau$ values appearing to the right and $\beta$ is the inverse
temperature ($\beta=1/T$).  We will work with the
Matsubara frequency Green's functions, defined for imaginary frequencies
$i\omega_n=i\pi T (2n+1)$.  The Green's function at each Matsubara frequency
is determined by a Fourier transformation
\begin{equation}
G_{ij}(i\omega_n)=\int_0^\beta d\tau e^{i\omega_n\tau}G_{ij}(\tau).
\label{eq: g_mats}
\end{equation}
We also will work with the analytic continuation of the time-ordered Green's
functions to the real axis (retarded or advanced Green's functions),
with $i\omega_n\rightarrow \omega\pm i0^+$.  We use the symbol $Z$ to denote
a general variable in the complex plane (although we will mainly be 
interested in either $Z=i\omega_n$ or $Z=\omega+i0^+$). Finally, we work in 
the mixed basis described above, where we Fourier transform the $x$- and
$y$-components to momentum space, to give $G_{\alpha\beta}({\bf k},Z)$,
where ${\bf R}_i$ has a $z$-component equal to $\alpha$ and ${\bf R}_j$
has a $z$-component equal to $\beta$ (${\bf k}$ is a two-dimensional
wavevector).

With all of this notation worked out, we can write the equation of motion
for the Green's function in real space\cite{potthoff_nolting_1999}, which 
satisfies
\begin{equation}
G_{ij}^{-1}(Z)=(Z+\mu)\delta_{ij}-\Sigma_{i}(Z)\delta_{ij}+t_{ij}.
\label{eq: g_eom_realspace}
\end{equation}
Now we go to a mixed-basis, by Fourier transforming in the $x$- 
and $y$-directions to find
\begin{eqnarray}
G_{\alpha\beta}^{-1}({\bf k},Z)&=&[Z+\mu-\Sigma_\alpha(Z)-
\epsilon^{2d}({\bf k})]\delta_{\alpha\beta}\nonumber\\
&+&t_{\alpha\alpha+1}\delta_{\alpha+1\beta}+t_{\alpha\alpha-1}
\delta_{\alpha-1\beta},
\label{eq: g_eom_mixed}
\end{eqnarray}
with $\Sigma_\alpha(Z)$ the local self energy for plane $\alpha$.
Finally, we use the identity $\sum_\gamma G_{\alpha\gamma}(Z)
G_{\gamma\beta}^{-1}(Z)=\delta_{\alpha\beta}$ to get the starting point for the 
recursive solution to the problem:
\begin{eqnarray}
\delta_{\alpha\beta}&=&
G_{\alpha\beta}({\bf k},Z)[Z+\mu-\Sigma_\beta(Z)-\epsilon_\beta^{2d}({\bf k})]
\nonumber\\
&+&G_{\alpha\beta-1}({\bf k},Z)t_{\beta-1\beta}+
G_{\alpha\beta+1}({\bf k},Z)t_{\beta+1\beta}.
\label{eq: eom1}
\end{eqnarray}
The equation 
of motion in Eq.~(\ref{eq: eom1}) has a tridiagonal form with respect to
the spatial component $z$, and hence it can be solved by employing the
renormalized perturbation expansion\cite{economou_1983}. We illustrate the 
solution exactly
here.  The equation with $\beta=\alpha$ is different from the equations
with $\beta\ne\alpha$.  The former is solved directly via
\begin{widetext}
\begin{equation}
G_{\alpha\alpha}({\bf k},Z)=\frac{1}
{Z+\mu-\Sigma_\alpha(Z)-\epsilon_\alpha^{2d}({\bf k})+
\frac{G_{\alpha\alpha-1}({\bf k},Z)}{G_{\alpha\alpha}({\bf k},Z)}
t_{\alpha-1\alpha}+
\frac{G_{\alpha\alpha+1}({\bf k},Z)}{G_{\alpha\alpha}({\bf k},Z)}
t_{\alpha+1\alpha}},
\label{eq: eom2}
\end{equation}
\end{widetext}
and the latter equations can all be put into the form
\begin{eqnarray}
-\frac{G_{\alpha\alpha-n+1}({\bf k},Z)t_{\alpha-n+1\alpha-n}}
{G_{\alpha\alpha-n}({\bf k},Z)}&=&Z+\mu-\Sigma_{\alpha-n}(Z)-
\epsilon_{\alpha-n}^{2d}({\bf k})\nonumber\\
&+&\frac{G_{\alpha\alpha-n-1}({\bf k},Z)t_{\alpha-n-1\alpha-n}}
{G_{\alpha\alpha-n}({\bf k},Z)},
\label{eq: eom3}
\end{eqnarray}
for $n>0$, with a similar result for the recurrence to the right.
We define the left function
\begin{equation}
L_{\alpha-n}({\bf k},Z)=
-\frac{G_{\alpha\alpha-n+1}({\bf k},Z)t_{\alpha-n+1\alpha-n}}
{G_{\alpha\alpha-n}({\bf k},Z)}
\label{eq: ldef}
\end{equation}
and then determine the recurrence relation from Eq.~(\ref{eq: eom3})
\begin{eqnarray}
L_{\alpha-n}({\bf k},Z)&=&Z+\mu-\Sigma_{\alpha-n}(Z)-
\epsilon_{\alpha-n}^{2d}({\bf k})\nonumber\\
&-&
\frac{t_{\alpha-n\alpha-n-1}t_{\alpha-n-1\alpha-n}}
{L_{\alpha-n-1}({\bf k},Z)}.
\label{eq: l_recurrence}
\end{eqnarray}
We solve the recurrence relation by starting with the result for
$L_{-\infty}$, and then iterating Eq.~(\ref{eq: l_recurrence}) up
to $n=1$.  Of course we do not actually go out infinitely far in
our calculations.  We assume we have semi-infinite metallic leads,
hence we can determine $L_{-\infty}$ by substituting $L_{-\infty}$
into both the left and right hand sides of Eq.~(\ref{eq: l_recurrence}),
which produces a quadratic equation for $L_{-\infty}$ that  is solved by
\begin{eqnarray}
L_{-\infty}({\bf k},Z)&=&
\frac{Z+\mu-\Sigma_{-\infty}(Z)-\epsilon_{-\infty}^{2d}({\bf k})}{2}
\label{eq: l_infty}\\
&\pm&\frac{1}{2}\sqrt{[Z+\mu-\Sigma_{-\infty}(Z)-
\epsilon_{-\infty}^{2d}({\bf k})]^2-4t_{-\infty}^2}.
\nonumber
\end{eqnarray}
The sign in Eq.~(\ref{eq: l_infty}) is chosen to yield an
imaginary part less than zero for $Z$ lying in the upper half plane,
and vice versa for $Z$ lying in the lower half plane.  If $L_{-\infty}$
is real, then we choose the root whose magnitude is larger than $t_{-\infty}$
(the product of the roots equals $t_{-\infty}^2$).  In our calculations,
we assume that the left function is equal to the value
$L_{-\infty}$ found in the bulk, until we are within thirty planes of the
first interface.  Then we allow those thirty planes to be self-consistently
determined with $L_{\alpha}$ possibly changing, and we include a similar
thirty planes on the right hand side of the last interface, terminating with 
the bulk result to the right as well.

In a similar fashion, we define a right function and a recurrence relation
to the right, with the right function
\begin{equation}
R_{\alpha+n}({\bf k},Z)=
-\frac{G_{\alpha\alpha+n-1}({\bf k},Z)t_{\alpha+n-1\alpha+n}}
{G_{\alpha\alpha+n}({\bf k},Z)}
\label{eq: rdef}
\end{equation}
and the recurrence relation
\begin{eqnarray}
R_{\alpha+n}({\bf k},Z)&=&Z+\mu-\Sigma_{\alpha+n}(Z)-
\epsilon_{\alpha+n}^{2d}({\bf k})\nonumber\\
&-&
\frac{t_{\alpha+n\alpha+n+1}t_{\alpha+n+1\alpha+n}}
{R_{\alpha+n+1}({\bf k},Z)}.
\label{eq: r_recurrence}
\end{eqnarray}
We solve the right recurrence relation by starting with the result for
$R_{\infty}$, and then iterating Eq.~(\ref{eq: r_recurrence}) up
to $n=1$.  As before,
we determine $R_{\infty}$ by substituting $R_{\infty}$
into both the left and right hand sides of Eq.~(\ref{eq: r_recurrence}),
which produces a quadratic equation for $R_{\infty}$ that  is solved by
\begin{eqnarray}
R_{\infty}({\bf k},Z)&=&
\frac{Z+\mu-\Sigma_{\infty}(Z)-\epsilon_{\infty}^{2d}({\bf k})}{2}
\label{eq: r_infty}\\
&\pm&\frac{1}{2}\sqrt{[Z+\mu-\Sigma_{\infty}(Z)-
\epsilon_{\infty}^{2d}({\bf k})]^2-4t_{\infty}^2}.
\nonumber
\end{eqnarray}
The sign in Eq.~(\ref{eq: r_infty}) is chosen the same way as for 
Eq.~(\ref{eq: l_infty}).  In our calculations,
we also assume that the right function is equal to the value
$R_{\infty}$ found in the bulk, until we are within thirty planes of the
first interface.  Then we allow those thirty planes to be self-consistently
determined with $R_{\alpha}$ possibly changing, and we include a similar
thirty planes on the left hand side of the last interface, terminating with
the bulk result to the left as well.

Using the right and left functions, we finally obtain the Green's function
\begin{equation}
G_{\alpha\alpha}({\bf k},Z)=\frac{1}{L_{\alpha}({\bf k},Z)+
R_{\alpha}({\bf k},Z)-[Z+\mu-\Sigma_\alpha(Z)-\epsilon_\alpha^{2d}({\bf k})]}
\label{eq: g_k_final}
\end{equation}
where we used Eqs.~(\ref{eq: l_recurrence}) and (\ref{eq: r_recurrence})
in Eq.~(\ref{eq: eom2}).  The local Green's function on each plane is then 
found by summing over the two-dimensional momenta, which can be replaced by an
integral over the two-dimensional density of states (DOS):
\begin{equation}
G_{\alpha\alpha}(Z)=\int d\epsilon^{2d}_\alpha \rho^{2d}(\epsilon^{2d}_\alpha)
G_{\alpha\alpha}(\epsilon^{2d}_\alpha,Z),
\label{eq: g_loc}
\end{equation}
with
\begin{equation}
\rho^{2d}(\epsilon^{2d}_\alpha)=\frac{1}{2\pi^2t_\alpha^\parallel a^2}
\mathbb{K}\left ( 1 - \sqrt{1-\frac{(\epsilon_\alpha^{2d})^2}
{(4t_\alpha^\parallel)^2}}\right ),
\label{eq: 2d_dos}
\end{equation}
and $\mathbb{K}(x)$ the complete elliptic integral of the first
kind.  If $t_\alpha^\parallel$
varies in the nanostructure, then changing variables to 
$\epsilon=\epsilon^{2d}_\alpha/t_\alpha^\parallel$ in Eq.~(\ref{eq: g_loc})
produces
\begin{equation}
G_{\alpha\alpha}(Z)=\int_{-4}^4 d\epsilon \frac{1}{2\pi^2a^2}
\mathbb{K}\left ( 1 - \sqrt{1- \frac{\epsilon^2}{16}}\right )
G_{\alpha\alpha}(t_\alpha^\parallel\epsilon,Z),
\label{eq: g_loc2}
\end{equation}
so that we can take the $\epsilon$ variable to run from $-4$ to $4$ for
the integration on every plane, and we just need to introduce the 
corresponding $t_\alpha^\parallel\epsilon$ substitution  
(for $\epsilon_\alpha^{2d}$) into the left and right recurrence relations.
In the bulk limit, where we use $t_\alpha=t$, we find that the local
Green's function found from Eqs.~(\ref{eq: g_loc}) and (\ref{eq: g_k_final})
reduce to the well-known expressions for the three-dimensional Green's
functions on a simple cubic lattice\cite{economou_1983}, with a hopping 
parameter $t$.

Once we have the local Green's function on each plane, we can perform the 
DMFT calculation to determine the local self
energy on each plane\cite{brandt_mielsch_1989,freericks_zlatic_2003b}.  
We start with Dyson's equation, which defines
the effective medium for each plane
\begin{equation}
G_{0\alpha}^{-1}(Z)=G_\alpha^{-1}(Z)+\Sigma_\alpha(Z).
\label{eq: dyson}
\end{equation}
The local Green's function for the $\alpha$th plane satisfies
\begin{equation}
G_\alpha(Z)=(1-w_1)\frac{1}{G_{0\alpha}^{-1}(Z)+\frac{1}{2}U}
+w_1\frac{1}{G_{0\alpha}^{-1}(Z)-\frac{1}{2}U},
\label{eq: impurity}
\end{equation}
with $w_1$ equal to the average filling of the localized particles
[note that this above form is slightly different from the usual 
notation\cite{freericks_zlatic_2003b}, 
because we have made the theory particle-hole symmetric by the choice of
the interaction in Eq.~(\ref{eq: hamiltonian}), so that $\mu=0$
corresponds to half filling in the barrier region and in the ballistic
metal leads]. Finally, the self energy is found from
\begin{equation}
\Sigma_\alpha(Z)=G_{0\alpha}^{-1}(Z)-G_\alpha^{-1}(Z).
\label{eq: sigma}
\end{equation}

The full dynamical mean field theory algorithm can now be stated.  We
begin by (i) making a choice for the self energy on each plane.  Next,
we (ii) use the left and right recurrences in Eqs.~(\ref{eq: l_recurrence}) and
(\ref{eq: r_recurrence}) along with the bulk values found in 
Eqs.~(\ref{eq: l_infty}) and (\ref{eq: r_infty}) and a choice for the
number of self-consistently determined planes within the metal leads
(which we choose to be 30 to the left and the right of the barrier
interfaces) to calculate the local Green's function at each plane
in the self-consistent region
from Eqs.~(\ref{eq: g_k_final}) and (\ref{eq: g_loc2}).  Once the local Green's
function is known for each plane, we then (iii) extract the effective medium
for each plane from Eq.~(\ref{eq: dyson}), (iv) determine the new local
Green's function from Eq.~(\ref{eq: impurity}), and (v) calculate the
new self energy on each plane from Eq.~(\ref{eq: sigma}). Then we iterate 
through steps (ii)--(v) until the calculations have converged.

For all of the calculations in this work, we will assume the hopping matrix is
unchanged in the metallic leads and the barrier, so all $t_{\alpha\alpha\pm 1}$ 
and all $t_\alpha^\parallel$ are equal to $t$, which we take as our
energy unit.  We also work at the particle-hole symmetric point of
half filling for the conduction electrons and the localized electrons.
This yields $w_1=1/2$ and $\mu=0$.

There are a number of numerical details that need to be discussed in these
computations.  First, one should note that the recurrence relations in
Eqs.~(\ref{eq: l_recurrence}) and (\ref{eq: r_recurrence}) always preserve
the imaginary part of $R$ or $L$ during the recursion.  Hence the recursion
is stable when $R$ or $L$ is complex.  On the other hand, when they are
real, we find that the large root is stable.  Since this is the physical
root, the recursion relations are always stable.  Second, the integrand can 
have a number of singularities in it. When we calculate the Matsubara
Green's functions, the only singularity comes from the logarithmic singularity
in the two-dimensional DOS.  We remove that singularity from the integration by
using a midpoint rectangular integration scheme for $0.5<|\epsilon|<4$,
and we change the variables for the region $|\epsilon|<0.5$ from
$\epsilon$ to $x^3=\epsilon$, which is finite as $x\rightarrow 0$, and
which has a finite slope as $x\rightarrow 0$; this allows a midpoint
rectangular integration scheme for $|x|<(0.5)^{1/3}$ to accurately determine
this second piece of the integral.  When we calculate the real frequency
Green's functions, we have the logarithmic singularity, but we also can have
a square-root singularity at the $\alpha$th plane in the denominator of the 
integrand when ${\rm Im}\Sigma_\alpha(\omega)=0$ and $|\omega+\mu-{\rm Re}
\Sigma_\alpha(\omega)-\epsilon|=2$.  We define 
$a=\omega+\mu-{\rm Re}\Sigma_\alpha(\omega)+2$ and
$b=\omega+\mu-{\rm Re}\Sigma_\alpha(\omega)-2$. Then, if $a<-4$ or $b>4$, the 
only singularity lies at $\epsilon=0$ as before.  When $b<-4$, but $-4<a<4$,
then there is a singularity at $\epsilon=a$; when $a>4$, but $-4<b<4$,
then there is a singularity at $\epsilon=b$; and when $-4<a,b<4$, there
are singularities at $a$ and $b$.  The singularities are easy to transform
away by using sine and hyperbolic cosine substitutions like
$\epsilon=\omega+\mu-{\rm Re}\Sigma_\alpha(\omega)-2\sin\theta$ and
$\epsilon=\omega+\mu-{\rm Re}\Sigma_\alpha(\omega)-2\cosh\theta$ into the
respective pieces of the integrands where a singularity lies.  We simply
determine where all possible singularities lie (for each plane), set up an 
appropriate
grid for the $\epsilon$ variable that takes the different changes of integration
variable into account, and compute the associated weight functions for the
integrations, in order to perform the integration over the two-dimensional
DOS.  Third, we find that when the correlations in the barrier are
strong enough that we are in the Mott insulator for the bulk material, and
the barrier is sufficiently thick, then the self energy develops a sharp
structure, where the real part goes through zero over a small range
close to $\omega=0$, and the imaginary part picks up a large delta-function-like
peak around $\omega=0$.  In order to properly pick up this behavior in the 
self-consistent solutions, we need to use a very fine integration grid
(we used up to one million points for the calculations reported on here)
to perform the integration over the two-dimensional DOS. Such a fine grid
is only needed for frequencies close to $\omega=0$, but one needs to have
a fine enough frequency grid in $\omega$ to pick up the sharp peak
behavior in the self energy (we use a step size of 0.001 when there is a sharp
structure in the self energy). For ordinary $\omega$ points, we typically used
an integration grid of 5000 points. Fourth, these equations are easy to
parallelize on the real-frequency axis, because the calculations for each value
of frequency are completely independent of one another, so we simply
use a master-slave approach and send the calculations at different frequencies
to each of the different slaves until all frequencies are calculated.  This
approach has an almost linear scale up in the parallelization speed.

In addition to these single-particle properties, we also are interested
in transport along the $z$-axis (perpendicular to the multilayered planes).
The resistance of the nanostructures can be calculated by a Kubo-based
linear response formalism\cite{kubo_1957} (i.e., a current-current correlation 
function).  We begin with the current operator at the $\alpha$th plane
\begin{eqnarray}
{\bf j}_z&=&\sum_\alpha {\bf j}_{z\alpha},\nonumber\\
{\bf j}_{z\alpha}&=&\frac{ieat}{\hbar}\sum_{i~{\rm in~2d~plane}}
\left ( c^\dagger_{\alpha i}c_{\alpha+1 i}-c^\dagger_{\alpha +1 i}c_{\alpha i}
\right ).
\label{eq: curr_def}
\end{eqnarray}
This operator sums all of the current flowing from the $\alpha$th plane to the
$\alpha+1$st plane.

The current-current correlation function is defined to be
\begin{equation}
\Pi_{\alpha\beta}(i\nu_l)=\int_0^\beta d\tau e^{i\nu_l\tau}\langle 
\mathcal{T}_\tau {\bf j}^\dagger_{z\alpha}(\tau){\bf j}_{z \beta}(0)\rangle,
\label{eq: curr-curr}
\end{equation}
with $i\nu_l=i\pi T 2l$ the Bosonic Matsubara frequency and with the dc
conductivity matrix determined by the analytic continuation of 
Eq.~(\ref{eq: curr-curr}) to the real frequency axis via
\begin{equation}
\sigma_{\alpha\beta}(\nu)=\lim_{\nu\rightarrow 0} {\rm Re}
\frac{i\hbar \Pi_{\alpha\beta}(\nu)}{\nu}.
\label{eq: sigma_def}
\end{equation}
Substituting Eq.~(\ref{eq: curr_def}) into Eq.~(\ref{eq: curr-curr}), 
evaluating the contractions in terms of the single-particle Green's functions,
performing the integration over $\tau$ to convert to the Matsubara
frequency representation, and performing a Fourier transform over the
2d-spatial coordinates, yields the following result after some
straightforward algebra:
\begin{eqnarray}
\Pi_{\alpha\beta}(i\nu_l)&=&\left ( \frac{eat}{\hbar}\right )^2
T\sum_m\sum_{\bf k}\Biggr \{\nonumber\\
&-&G_{\beta+1\alpha+1}({\bf k},i\omega_m)G_{\alpha\beta}({\bf k},
i\omega_m+i\nu_l)\nonumber\\
&+&G_{\beta\alpha +1}({\bf k},i\omega_m)G_{\alpha\beta+1}({\bf k},
i\omega_m+i\nu_l)\nonumber\\
&+&G_{\beta+1\alpha}({\bf k},i\omega_m)G_{\alpha+1\beta}({\bf k},
i\omega_m+i\nu_l)\nonumber\\
&-&G_{\beta\alpha}({\bf k},i\omega_m)G_{\alpha+1\beta+1}({\bf k},
i\omega_m+i\nu_l)\Biggr \} .
\label{eq: pi_mats}
\end{eqnarray}
Now we need to perform the analytic continuation from the imaginary to the
real frequency axis\cite{mahan_1990}.  
This is done by first converting the summations
over the Matsubara frequencies into contour integrals that enclose
all of the Matsubara frequencies and are multiplied by the Fermi-Dirac 
distribution function $f(\omega)=1/[1+\exp(\beta\omega)]$
which has a pole at each Matsubara frequency.  Then the contours are
deformed to go along lines parallel (but just above or just below) the
real axis, and the real axis shifted by $-i\nu_l$. At this point we 
replace $f(\omega-i\nu_l)$ by $f(\omega)$ and then analytically continue
$i\nu_l\rightarrow \nu+i0^+$.  The algebra is once again straightforward
but somewhat tedious.  The final result is
\begin{eqnarray}
\Pi_{\alpha\beta}(\nu)&=&-\frac{1}{\pi}\left (\frac{eat}{\hbar}\right )^2
\sum_{\bf k}\Biggr [ f(\omega)\Big \{\nonumber\\
&~&G_{\alpha\beta}({\bf k},\omega+\nu){\rm Im}G_{\beta+1\alpha+1}({\bf k},
\omega)\nonumber\\
&+&G_{\alpha\beta+1}({\bf k},\omega+\nu){\rm Im}G_{\beta\alpha+1}({\bf k},
\omega)\nonumber\\
&+&G_{\alpha+1\beta}({\bf k},\omega+\nu){\rm Im}G_{\beta+1\alpha}({\bf k},
\omega)\nonumber\\
&-&G_{\alpha+1\beta+1}({\bf k},\omega+\nu){\rm Im}G_{\beta\alpha}({\bf k},
\omega)\Big \}\nonumber\\
&+&f(\omega+\nu)\Big \{\nonumber\\
&-&G^*_{\beta+1\alpha+1}({\bf k},\omega){\rm Im}G_{\alpha
\beta}({\bf k},\omega+\nu)\nonumber\\
&+&G^*_{\beta\alpha+1}({\bf k},\omega){\rm Im}G_{\alpha\beta+1}({\bf k},
\omega+\nu)\nonumber\\
&+&G^*_{\beta+1\alpha}({\bf k},\omega){\rm Im}G_{\alpha+1\beta}({\bf k},
\omega+\nu)\nonumber\\
&-&G^*_{\beta\alpha}({\bf k},\omega){\rm Im}G_{\alpha+1\beta+1}({\bf k},
\omega+\nu)\Big \} \Biggr ].
\label{eq: pi_final}
\end{eqnarray}
The last step is to evaluate the dc conductivity matrix, which becomes
\begin{eqnarray}
\sigma_{\alpha\beta}(0)&=&\frac{2e^2}{h}a^2t^2\int d\epsilon \rho^{2d}(\epsilon)
\int d\omega \left ( -\frac{d f}{d\omega}\right )\nonumber\\
&\Big [ & {\rm Im}G_{\beta\alpha+1}(\epsilon,\omega){\rm Im}G_{\alpha\beta+1}
(\epsilon,\omega)\nonumber\\
&+&{\rm Im}G_{\beta+1\alpha}(\epsilon,\omega)
{\rm Im}G_{\alpha+1\beta}(\epsilon,\omega)\nonumber\\
&-&{\rm Im}G_{\beta+1\alpha+1}(\epsilon,\omega)
{\rm Im}G_{\alpha\beta}(\epsilon,\omega)\nonumber\\
&+&{\rm Im}G_{\beta\alpha}(\epsilon,\omega)
{\rm Im}G_{\alpha+1\beta+1}(\epsilon,\omega)\Big ] . 
\label{eq: sigmadc_final}
\end{eqnarray}
The conductivity matrix has the dimensions $e^2/ha^2$, which is the inverse
of the resistance unit, divided by two factors of length, and is the correct
units for the conductivity matrix.

Since the conductivity matrix is not as familiar as the scalar conductivity
used for homogeneous problems, we will briefly derive how one extracts
the resistance of the nanostructure from the conductivity matrix.  The 
key element that we use is that the current density
that flows through each plane
is conserved, because charge current can neither be created nor destroyed
in our device.  The continuity equation, then says that the current
density through the $\alpha$th plane, $I_\alpha$, is related to the electric 
field, $E_\beta$, between the $\beta$th and $\beta+1$st plane via
\begin{equation}
I_\alpha=a\sum_\beta \sigma_{\alpha\beta}(0)E_\beta=I,
\label{eq: cont}
\end{equation}
where we set the current density on each plane equal to a constant value $I$.
Inverting this relation to determine the electric field gives
\begin{equation}
E_\beta=\frac{1}{a}\sum_\alpha [\sigma^{-1}(0)]_{\beta\alpha}I.
\label{eq: e_solver}
\end{equation}
The voltage across the nanostructure is just the sum of the electric field
between each plane, multiplied by the interplane distance (we assume a
constant dielectric constant throughout), so we can 
immediately determine the resistance-area product (specific resistance)
from Ohm's law
\begin{equation}
R_na^2=\frac{V}{I}=\sum_{\alpha\beta}[\sigma^{-1}(0)]_{\beta\alpha}.
\label{eq: resistance}
\end{equation}
One needs to pursue a similar type of analysis to examine the thermal
transport properties (thermopower and thermal resistance), but it is
somewhat more complicated, because the thermal current is not conserved
from one plane to another plane, as is the charge current. We will present 
results for such a calculation elsewhere (at half filling, where we restrict 
ourselves in this paper, there is no thermopower by particle-hole
symmetry).

The only mathematical issue associated with this analysis is that we have
assumed the conductivity matrix is invertible.  In general, this is not true
when there is no scattering in the metallic leads.  In this case, we
need to truncate the conductivity matrix to consider only the block that
covers all of the planes in the barrier and the first metallic plane to the left
and to the right of the barrier.  This matrix is always invertible, 
and allows calculations to be performed easily (if we were to include 
a larger matrix, we find that the resistance does not increase as we increase
the number of planes within the metallic leads that we include in the
conductivity matrix block that is inverted, at least until we run into 
precision issues for the calculations).  Of course, if the metallic leads
have scattering, there are no numerical issues associated with the matrix 
inversion (except when the matrix is made too large and the system has 
approached the bulk limit, see below), but we need to decide how far down
the metallic leads we will perform the actual measurement, since the
voltage grows with the thickness of the metallic leads included in the
calculation (when there is scattering in the leads).

In order to calculate the dc conductivity matrix in 
Eq.~(\ref{eq: sigmadc_final}), we need to evaluate the off-diagonal components 
of the Green's functions.  This is easy to do using the renormalized
perturbation expansion, and the right and left functions.  We find two
recurrence relations
\begin{equation}
G_{\alpha\alpha-n}(\epsilon,\omega)=-\frac{G_{\alpha\alpha-n+1}
t_{\alpha-n+1\alpha-n}}{L_{\alpha-n}(\epsilon,\omega)},
\label{eq: g_od_lrecurrence}
\end{equation}
(defined for $n>0$) and
\begin{equation}
G_{\alpha\alpha+n}(\epsilon,\omega)=-\frac{G_{\alpha\alpha+n-1}
t_{\alpha+n-1\alpha+n}}{R_{\alpha+n}(\epsilon,\omega)},
\label{eq: g_od_rrecurrence}
\end{equation}
(also defined for $n>0$). The other off-diagonal Green's functions are
found from the symmetry relations: $G_{\alpha\alpha-n}=G_{\alpha-n\alpha}$
and $G_{\alpha\alpha+n}=G_{\alpha+n\alpha}$.

The computation of the junction resistance for a given temperature is
relatively simple to perform.  First, one must calculate all of
the local self energies for each plane, using the algorithm described
above.  Then, for each frequency $\omega$, one can calculate all of the
Green's functions that enter into the formula for $\sigma_{\alpha\beta}(0)$.
It is best to evaluate the integral over $\omega$ for many different 
temperatures ``at the same time'' since the only thing that changes
with temperature (when at half filling, where the chemical potential is
fixed and does not vary with $T$) is the Fermi factor derivative.  Since
evaluating at each frequency is independent of every other frequency,
this algorithm is also ``embarrassingly parallel''.

One final comment is in order about the formalism for calculating the
junction resistance. Namely, how does it relate to a Landauer approach
to the resistance?  In the Landauer approach\cite{datta_1995} one does not 
calculate a conductivity matrix, but instead determines the transport directly 
by evaluating the Green's function $G_{\alpha\beta}$ where $\alpha$ lies
at the left interface and $\beta$ lies at the right interface. We 
believe one can show that these two approaches are completely equivalent if one
uses the same self energies for the inhomogeneous structure to calculate
the Green's functions that enter into the transport calculation.  We
will examine this relationship in a future publication.

In a homogeneous (bulk) noninteracting 
system, we find that the Green's functions satisfy
\begin{eqnarray}
G_{\alpha\alpha\pm n}(\epsilon,\omega)&=&\frac{-i}{\sqrt{4t^2-(\omega+\mu
-\epsilon)^2}}\label{eq: g_od_bulk}\\
&\times&\left [ -\frac{\omega+\mu-\epsilon}{2}+i\frac{\sqrt{4t^2-
(\omega+\mu-\epsilon)^2}}{2}\right ]^n\nonumber
\end{eqnarray}
when $\epsilon$ lies within the band [$|\omega+\mu-\epsilon|<2$].  
Note that ${\rm Im}G_{\alpha\beta}
(\epsilon,\omega)$ is not always negative when $\alpha\ne\beta$.  This
occurs because we are using a mixed basis, and the imaginary part of the
Green's function does not have a definite sign in this basis. We can
substitute these Green's functions into the expression for the
conductivity matrix, to evaluate the result for the bulk.  We find
the matrix has all of its matrix elements equal to each other, and
they assume the value
\begin{equation}
\sigma_{\alpha\beta}(0)=\frac{e^2}{ha^2}\int_{-2}^2 d\epsilon \rho^{2d}
(\epsilon)\approx 0.63 \frac{e^2}{ha^2},
\label{eq: sharvin}
\end{equation}
for the case of half filling $\mu=0$ (since every matrix element is the
same, the conductivity matrix is not invertible, but the resistance
can still be calculated).  This result will lead to precisely the
Sharvin contact 
resistance\cite{sharvin_1965_russia,sharvin_1965,nikolic_freericks_miller_2001} 
when we convert the conductivity into a resistance
(the resistivity of a ballistic metal vanishes, but the resistance is nonzero).

\section{Single-particle properties}

We perform our calculations at half filling ($\mu=0$, $\langle c^\dagger_ic_i
\rangle=1/2$, and $w_1=\langle w_i \rangle=1/2$).  This has a number
of advantages.  First, because the chemical potential is the same for the
metallic leads and the barrier, there is no electrochemical force that
reorganizes the electrons to a screened dipole layer at each of the interfaces,
instead the filling remains homogeneous throughout the system.  Second, 
the chemical potential is fixed as a function of temperature, so there
is no need to perform imaginary-axis calculations to determine the
chemical potential as a function of temperature.  We usually calculate the
Matsubara Green's functions anyway, to test the accuracy of the real-axis
Green's function, by comparing the Matsubara Green's functions calculated
directly with those calculated from the spectral formula via the 
real-axis DOS (usually the accuracy is better than three decimal points
for every Matsubara frequency).  Third, we can perform calculations of
the resistance at all temperatures in parallel, because the chemical
potential does not vary with temperature (recall, the DOS of the Falicov-Kimball
model is temperature independent for the DMFT solution\cite{vandongen_1992}).  
Fourth, the particle-hole
symmetry of the DOS allows us to have another check on the accuracy of
the calculations because we do not invoke that symmetry in our calculations.
Fifth, there is a metal-insulator transition (MIT) in the bulk Falicov-Kimball
model on a cubic lattice when $U\approx 4.9t$, so the solutions at
half filling include the MIT.
For these reasons, we find this case to be the simplest one to consider
in a first approach to the inhomogeneous many-body problem.

\begin{figure}[htbf]
\epsfxsize=3.0in
\epsffile{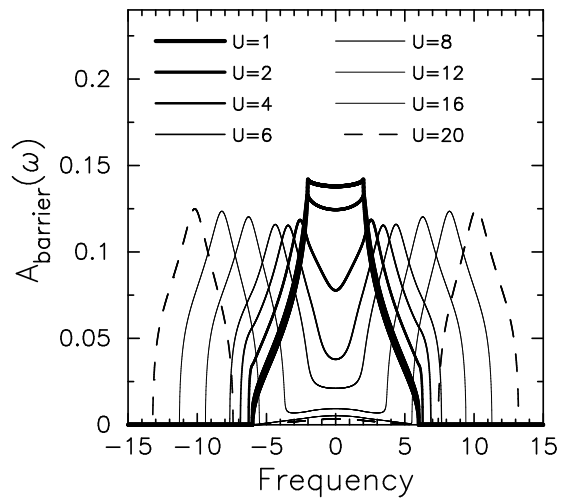}
\caption{\label{fig: n=1} Barrier DOS as a function of the Falicov-Kimball
interaction $U$.  The different line widths and styles denote different
$U$ values, as detailed in the legend.  Note how the DOS initially evolves
as in the bulk, with the DOS being reduced near $\omega=0$, and the
band width increasing.  But as we pass through the Mott transition, we see
that the double-peak Mott-Hubbard bands appear, but so does a low-energy
(interface-localized) band near $\omega=0$, which looks like a low-weight
metallic band for large $U$.
}
\end{figure}

\begin{figure}[htbf]
\epsfxsize=3.0in
\epsffile{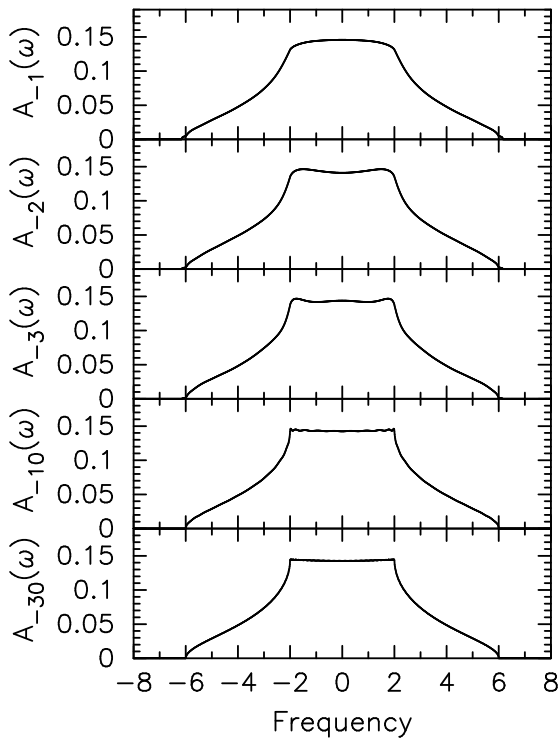}
\caption{\label{fig: u=2_lead} Lead DOS for an $N=5$ barrier device
with $U=2$. The different panels show the DOS in the first metal plane
to the left of the barrier, in the second, the third, the tenth and the
thirtieth.  Note how the system approaches the bulk cubic DOS as it moves
further from the interface, as expected.  A careful examination of the
panels shows that the ``flat'' region with $|\omega|<2$ shows a
half-period oscillation for each unit of distance from the current plane to the
interface, but the amplitude shrinks dramatically as we move further
from the interface.
}
\end{figure}

\begin{figure}[htbf]
\epsfxsize=3.0in
\epsffile{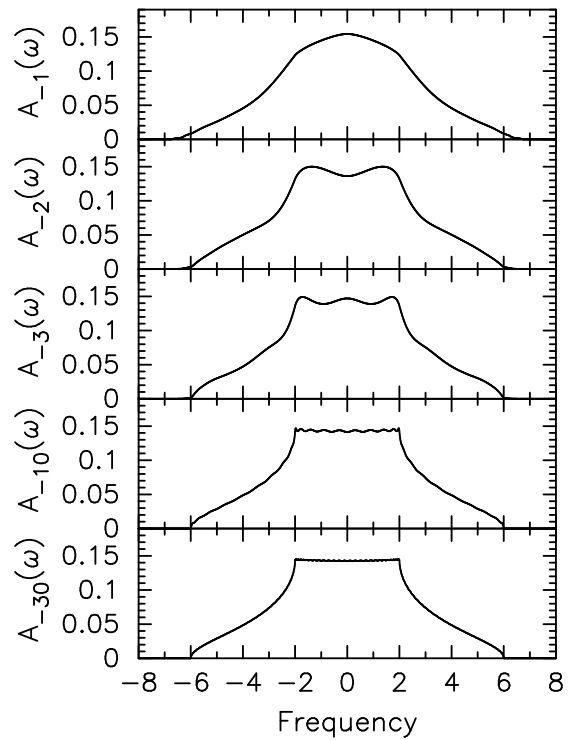}
\caption{\label{fig: u=4_lead} Lead DOS for an $N=5$ barrier device
with $U=4$. The different panels show the DOS in the first metal plane
to the left of the barrier, in the second, the third, the tenth and the
thirtieth.   Note how the amplitude of the oscillations increases as
$U$ increases.
}
\end{figure}

\begin{figure}[htbf]
\epsfxsize=3.0in
\epsffile{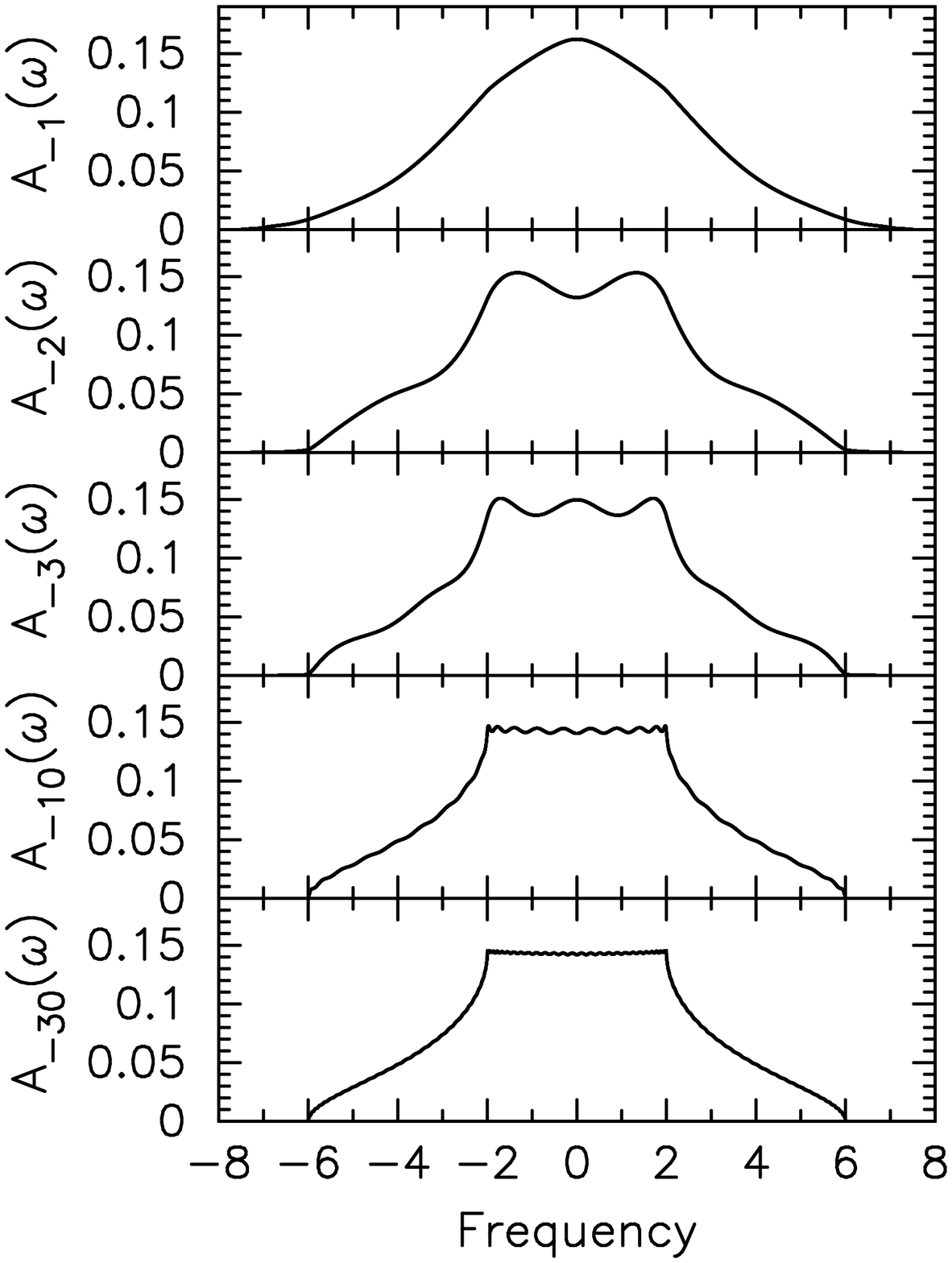}
\caption{\label{fig: u=6_lead} Lead DOS for an $N=5$ barrier device
with $U=6$. The different panels show the DOS in the first metal plane
to the left of the barrier, in the second, the third, the tenth and the
thirtieth.  Note how the amplitude of the oscillations is even larger
here.  A careful examination shows there are also oscillations (with
the same kind of increase in the number of half periods with the
distance from the interface) in the region $|\omega  |>2$.
}
\end{figure}

We also reduce the number of parameters in our calculations by assuming
all of the hopping matrix elements are equal to $t$ for nearest neighbors.
This is by no means necessary, but it allows us to reduce the number of
parameters that we vary in our calculations, which allows us to focus
on the physical properties with fewer calculations.  The hopping 
scale $t$ is used as our energy scale.  We also include 30 self-consistent
planes in the metallic leads to the left and to the right of our barrier,
which is varied between 1 and 80 planes in our calculations.

The first problem we investigate is the extreme quantum limit of having one
atomic plane in the barrier of our device.  We tune the Falicov-Kimball
interaction in the one barrier plane from $U=1$ to $U=20$, which goes from
a dirty metal to well into the Mott insulating regime.  But the Mott
insulating phase does not like being confined to a single atomic plane,
and there is a metallic proximity effect, where the metallic DOS leaks into
the insulator DOS at low energies.  The result is that we do not
expect the single-plane barrier to be too resistive. This is easiest to 
see when we consider the local DOS within the barrier plane, as
plotted in Fig.~\ref{fig: n=1}.  There we see that the DOS starts to be
reduced at the chemical potential as we in crease $U$, but there is still
substantial DOS at the Fermi energy when $U\approx 4.9$.  In fact, as
$U$ is increased, we see that the upper and lower Mott-Hubbard bands form,
centered at $\pm U/2$, but there is significant DOS that remains centered
at $\omega=0$, and it even develops a small peak for $U>10$. The origin of,
and the size of this peak, can be shown to arise naturally from the
renormalized perturbation theory expressions for the Green's
functions, but we do not do so here\cite{freericks_unpub}. We anticipate that 
these states
are localized at the interface, and represent the states that an incident
electron can tunnel through to go from one metallic lead to the other
in a transport experiment.  These results show a number of interesting
features of the coupling of a Mott insulator to a metallic lead: (i)
the Mott transition remains in the sense that Mott-Hubbard bands
continue to form, with their origin clearly seen near the MIT; (ii)
the interface-localized states have a metallic character (i.e., a peak
at $\omega=0$) in the large-$U$ regime; and
(iii) the proximity effect appears to
always be active, and able to create states within the barrier at low energy,
but the total weight in those states is low, so medium to high energy
properties of the Mott insulator phase will remain similar to the bulk.

Next we examine what happens as we increase the barrier thickness for given
values of $U$.  Our focus is on three generic values of interest: $U=2$,
which is a strongly scattering, diffusive metal; $U=4$, which is so close to 
the MIT, that the bulk DOS show a significant dip near $\omega=0$; and
$U=6$, which is well within the Mott-insulating phase.  We first examine how the
metallic leads are influenced by the presence of the barrier.  We set the origin
of the $\alpha$ variables so that $\alpha=0$ corresponds to the first barrier
plane (hence planes $-1$ to $-30$ represent the thirty planes to the
left of the barrier, with $-1$ closest to the barrier).  In 
Fig.~\ref{fig: u=2_lead}, we show results for $U=2$ and five representative
planes in the metal (the device has five barrier planes).  
In Fig.~\ref{fig: u=4_lead}, we show the same results for $U=4$ and in
Fig.~\ref{fig: u=6_lead}, we show the same results for $U=6$.
The first thing to notice is that the DOS is close to that of the bulk
simple cubic lattice for 30 planes away from the interface, indicating that
our choice of thirty self-consistent planes is reasonable.  Next, note that
the amplitude of the oscillations grows as $U$ increases.  Third, the number
of half periods in the oscillations increases with the distance away from
the interface (both for $|\omega|<2$ and $|\omega|>2$).  The source of these
oscillations is the Friedel oscillations (with a wavelength on the order
of two lattice spacings for half filling) that we expect associated with the
disturbance of the Fermi sea of the metal by the proximity to the
interface.

There are two interesting questions to ask about these results:
how thick does the barrier have to be before the Friedel oscillations
become frozen in the metallic leads and don't change with a thicker 
barrier, and do we see oscillatory behavior in the barrier, where we instead 
expect there to be exponentially decaying wavefunctions?  We find that the
answer to the first question is that the structure is already essentially
frozen in for a single-plane barrier, and it does not evolve much
with the barrier thickness (although it does show much evolution with
the interaction strength).  This perhaps sheds some light on why 
non-self-consistent Landauer based approaches for transport have been
so successful.  If one has a good guess for the semi-infinite lead
DOS, then it does not change much as the thickness increases, so that
guess will work well for all calculations with the same strength of
electron correlations.

\begin{figure}[htbf]
\epsfxsize=3.0in
\epsffile{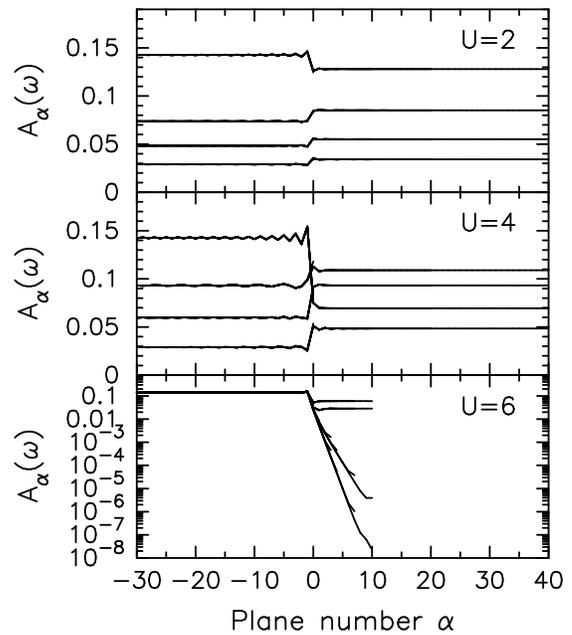}
\caption{\label{fig: z-slice} DOS at specific values of $\omega$ as a
function of the plane position in the device.  We plot only the left-hand
piece of the plots, since the right-hand piece is a mirror image of the
left-hand piece. Note that the $U=6$ panel is a semilogarithmic plot.
The four values of $\omega$ for $U=2$ are 0.0, 3.0, 4.0, and 5.0.  The
barrier thicknesses are $N=1$, 5, 10, 20, 40, and 80.  The four values
of $\omega$ for $U=4$ are 0.0, 2.5, 3.5, and 5.0.  The barrier thicknesses
are $N=1$, 5, 10, 20, 40, and 80.  The four values of $\omega$ for $U=6$
are 0.0, 0.2, 0.4, and 1.0.  The thicknesses are $N=1$, 4, 7, 10, 15,
and 20.  Note how all curves lie on top of each other in the metallic lead,
indicating the structure in the metallic lead is frozen in 
for an $N=1$ barrier, and does not significantly change with 
increasing $N$.  In the barrier, we
only have oscillations at the interface, and then the curves either are
flat with thickness ($U=2$ and 4), or exponentially decreasing or flat
($U=6$).  The little tails that stick out for the lowest two frequencies
with $U=6$ show that the middle plane of the barrier does not follow the
same exponential decay as the other planes do.  But the exponent of the
exponential decay is frozen in starting at $N\approx 1$.
}
\end{figure}

To examine the second question, we plot results for the DOS at a fixed
frequency (four chosen for each $U$ value) in Fig.~\ref{fig: z-slice}.
There are six different thicknesses plotted for each $U$ value.
The curves all lie on top of 
each other for the metallic lead planes, indicating that the Friedel oscillation
structure is frozen in starting at $N=1$ (and we can read off the oscillation
wavelength to be two lattice spacings, with a sharp decrease of the 
amplitude as one moves away from the interface).  In the barrier, we see
that there are only oscillations close to the interface, then the curves
either flatten out or exponentially decay with thickness.  But the curves
continue to lie on top of each other (except for the middle plane of the
barrier for small $\omega$ and $U=6$).  These results, once again, show that 
another of the assumptions of the non-self-consistent Landauer-based
approaches, that there is an exponential decay with a well defined 
decay length in the insulating barrier regions, holds here as well, but
one needs to properly predict the decay length to perform accurate calculations.



\begin{figure}[htbf]
\epsfxsize=3.0in
\epsffile{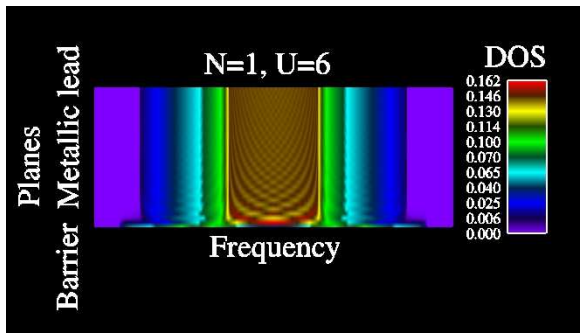}
\caption{\label{fig: n=1_u=6} False-color plot of the DOS for a $N=1$
barrier plane device with $U=6$. The barrier plane is just the lowest plane
at the bottom of the figure, while the thirty metallic planes lie on
top. Note how the ripples of the Friedel oscillations are most visible
in the central region, where the DOS has a plateau.  (Color version online.)
}
\end{figure}

\begin{figure}[htbf]
\epsfxsize=3.0in
\epsffile{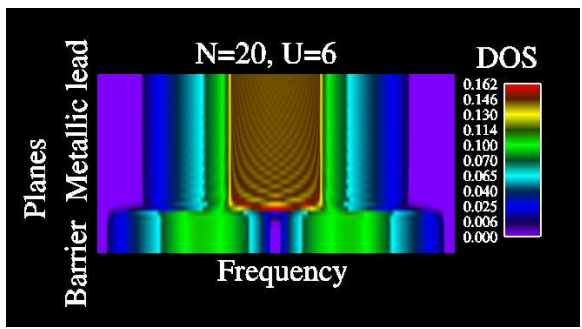}
\caption{\label{fig: n=20_u=6} False-color plot of the DOS for a $N=20$
barrier plane device with $U=6$. The barrier planes are  the lower  ten planes,
while the thirty metallic planes lie on
top. Note how the ripples of the Friedel oscillations agree with those
in Fig.~\ref{fig: n=1_u=6}. In the barrier, the DOS decreases rapidly on this
linear scale, and shows few oscillations, but one can see some small 
oscillations near the band edges in both regions.  (Color version online.)
}
\end{figure}

Our final summary of the DOS is included in false color plots (the color,
or grayscale, denoting the height of the DOS at a given plane) to emphasize
the spatial location and amplitudes in the oscillations.  
Fig.~\ref{fig: n=1_u=6} shows the results for $N=1$ and $U=6$ and
Fig.~\ref{fig: n=20_u=6} shows the results with $N=20$ and $U=6$
(only half of the nanostructure planes are shown due to the mirror symmetry).
The color scale (or grayscale) needs to use a banded rainbow, with the
different colors (grayscales) separated by bands of black in order to
pick up the small amplitude oscillations in the background of the large
DOS.  Note how the Friedel oscillations are essentially the same in the
two plots, indicating this freezing of the oscillations starting at
$N=1$.  There are also oscillations visible near the metal band edges,
indicating Friedel-like oscillations due to the different total bandwidths
of the two materials joined in the nanostructure.  The DOS in the barrier at
low frequency becomes very small very quickly on these linear scales,
but it is nonzero (see Fig.~\ref{fig: z-slice}).

\begin{figure}[htbf]
\epsfxsize=3.0in
\epsffile{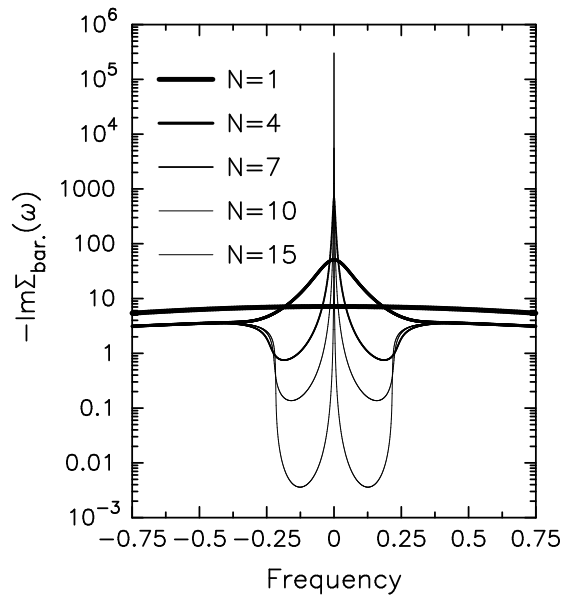}
\caption{\label{fig: self-energy} 
Semilogarithmic plot of the imaginary part of the self energy on the central
plane of the barrier at small frequency for five different thickness
barriers ($N=1$, 4, 7, 10, and 15). Note how the imaginary part of the
self energy becomes very small for frequencies close to $\omega=0$, but
as we approach $\omega=0$, a sharp delta-function-like peak develops
that narrows as the barrier is made thicker.  It is precisely this structure
that is hard to reproduce with numerical calculations.  Note that this kind
of a self energy is very similar to what is seen in the hypercubic
lattice in infinite dimensions.
}
\end{figure}

The final single-particle property we consider is the imaginary part of the
self energy at the central plane of the barrier at low energy in
Fig.~\ref{fig: self-energy}. In the bulk,
the imaginary part of the self energy vanishes within the Mott-Hubbard gap,
except for a delta function at $\omega=0$ whose weight can be used
as a quasi-order parameter for the Mott transition at half filling
(but not away from half filling\cite{demchenko_joura_freericks_2004}). 
In the nanostructures, the imaginary
part of the self energy never vanishes in the bulk gap region, but it can
assume very small values, with a sharp peak, of finite width, developing
at $\omega=0$.  This peak grows in height and narrows as the barrier is
made thicker.  It is a challenge to try to calculate such a structure 
numerically, especially due to the loss of precision in extracting the
self energy from the Dyson equation during the iterative algorithm.  It
requires a fine enough frequency grid to pick up the narrow structure,
and it requires a sufficiently fine integration grid for $\epsilon$,
in order to accurately determine the peak value.  Note how the self energy
evolves from a relatively broad featureless structure to a very sharply peaked
structure as the barrier is made thicker. This kind of a peaked self energy
is similar to what is seen in the exact solution on the hypercubic lattice
in infinite dimensions.  There the Mott transition is actually to a 
pseudogap phase, with the DOS vanishing only at the chemical potential,
but there is a region of exponentially small DOS in the ``gap region''.
The sharp features in the self energy led to a significant enhancement of the
low-temperature thermopower on the hypercubic lattice, when the system was
doped off of half filling\cite{freericks_etal_2003}
(and $w_1$ changed to produce an insulator). It is
unclear at this point whether such behavior could lead to enhancements in the
nanostructures, even though the self energy has similar properties.

\section{Generalized Thouless energy}

It is important to try to bring semiclassical ideas of transport into
transport in nanostructures, to see whether those concepts have useful
quantum analogues.  Thouless was the first to investigate such ideas for
diffusive metal barriers\cite{edwards_thouless_1972,thouless_1974}.  
He considered the idea of a dwell time in the
barrier for an electron that tries to travel through the barrier.  If 
we assume the electron takes a random walk through the barrier, then 
the time it spends inside the barrier is proportional to the
square of the thickness of the barrier (with the proportionality being
related to the diffusion constant).  Since one can extract the diffusion
constant, via an Einstein relation, from the junction resistance, Thouless
could construct a quantum-mechanical energy $\hbar/t_{dwell}$ from these
classical ideas.  It turns out that this energy scale plays a significant
role in determining the quantum dynamics of many different
kinds of nanostructures.  For example, it can be easily generalized to
take into account ballistic metals, where $t_{dwell}=Na/v_F$ for a barrier
of thickness $Na$, with $v_F$ the Fermi velocity.
The Thouless energy appears to be the critical quantum energy scale that
determines the dynamics through weakly correlated nanostructures; its
success in the theory of Josephson junctions is particularly 
noteworthy\cite{dubos_etal_2001}.

\begin{figure}[htbf]
\epsfxsize=3.0in
\epsffile{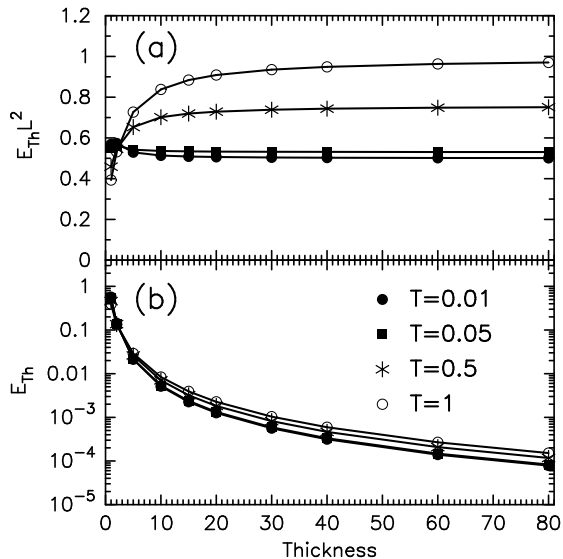}
\caption{\label{fig: thouless_u=4}
Thouless energy for a $U=4$ (diffusive, but very strongly scattering metal)
barrier as a function of the barrier thickness $L=Na$.  The different curves
correspond to different temperatures.  The top panel multiplies the Thouless
energy by $L^2$ to try to isolate the prefactor for the diffusive transport,
while the bottom panel plots the Thouless energy on a semi-logarithmic plot.
Note that the temperature dependence of the constant, seen for thick barriers
in panel (a), arises from the fact that the $U=4$ DOS has significant low-energy
structure, because there is a dip that develops near the chemical potential,
so the temperature dependence is both stronger than expected for normal
metals, and anomalous because many more states are involved as $T$ is
increased, i.e. it behaves more like an insulator.
}
\end{figure}

So the fundamental question we wish to investigate is can the concept of
a Thouless energy be generalized to a strongly correlated system, where
transport through a nanostructure is either via tunneling or via
incoherent thermal excitation.  The answer is yes, and we do so by first
trying to extract an energy scale from the resistance of the junction,
which is able to track the putative thermal dependence of the resistance
when we are in the incoherent thermal transport regime.  A simple dimensionality
argument shows that the form
\begin{equation}
E_{Th}=\frac{\hbar}{R_na^22e^2\int d\omega [-df/d\omega] \rho_{bulk}(\omega) Na}
\label{eq: thouless}
\end{equation}
has the the kind of dependence we are looking for.  The symbol 
$\rho_{bulk}(\omega)$ is the local DOS in the bulk for the material
that sits in the barrier of the nanostructure. If we check the dimensions,
we see that $R_n$ has dimensions $h/e^2$, and the DOS has dimensions $1/a^3t$,
so $E_{Th}$ is an energy [note Eq.~(\ref{eq: thouless}) corrects 
typos in an earlier work\cite{freericks_2004}].  When we examine systems 
where the barrier is 
a metal, then at low temperature the bulk DOS can be replaced by a constant
in the integral, and we reproduce the known forms for the Thouless energy
for ballistic ($E_{Th}\approx C/Na$) and diffusive ($E_{Th}\approx 
C^\prime/[Na]^2$) electrons
because the resistance is independent of the thickness for 
a ballistic metal barrier and it grows linearly with the thickness for a 
diffusive metal barrier.  This method of generalizing the Thouless energy
also avoids us having to try to answer the question of how long does it
take an electron to tunnel from the left to the right lead, and it reproduces 
all of the known forms for the Thouless energy in a unifying formula that
does not require us to even use the Einstein relation to extract a
diffusion constant or to determine the Fermi velocity for an anisotropic
Fermi surface (in the ballistic case).

We plot the results for this Thouless energy as a function of thickness
in Fig.~\ref{fig: thouless_u=4} for $U=4$.  In panel (a), we multiply
$E_{Th}$ by the square of the length $L=Na$ of the barrier. The different
curves correspond to different temperatures.  If the Thouless energy went 
exactly like $C^\prime/L^2$, then all of the curves would be straight lines,
with a temperature-dependent value $C^\prime(T)$.  But we see some curvature
for small barrier thicknesses.  This arises mainly from the fact that in
addition to the diffusive contribution to the resistance, there is a 
contact resistance, so for thin barriers, we do not have a pure 
$1/L^2$ behavior.  Note, however, that the Thouless energy has little
temperature dependence at low temperature, as expected. In panel (b), we
plot the curves on a semi-logarithmic plot, so one can see how
small the Thouless energy becomes for thicker junctions.

\begin{figure}[htbf]
\epsfxsize=3.0in
\epsffile{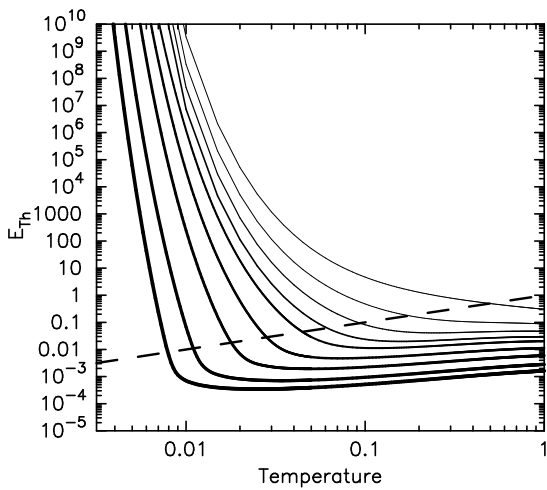}
\caption{\label{fig: thouless_u=6}
Thouless energy for a $U=6$ (Mott-insulating)
barrier as a function of temperature on a log-log plot.  The different curves
correspond to different thicknesses of the barrier, ranging from $N=1$ for
the top curve to $N=2$, 3, 4, 5, 7, 10, 15, and 20 as we move down the
plot.  Note how the Thouless energy picks up dramatic temperature dependence
here.  The dashed line is the curve where $E_{Th}=T$. We find that when the
Thouless energy equals the temperature, interesting effects occur (see below).
}
\end{figure}

\begin{figure}[htbf]
\epsfxsize=3.0in
\epsffile{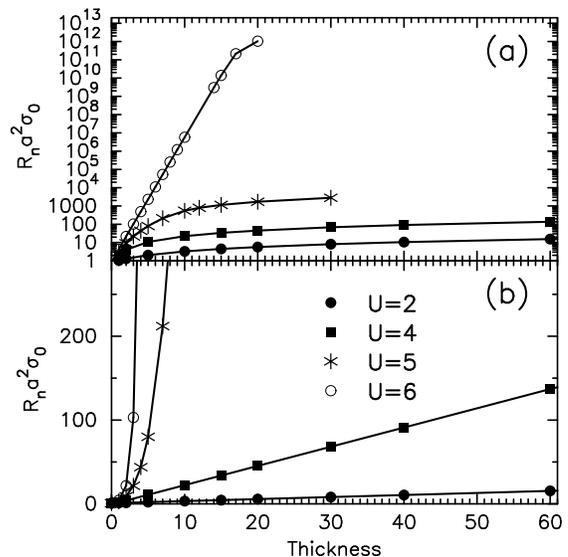}
\caption{\label{fig: rn_thick}
Resistance-area product for nanostructures with $U=2$, 4, 5, and 6, and
various thicknesses. Panel (a) is a semi-logarithmic plot, while panel (b)
is a linear plot.  The temperature is $T=0.01$ in both panels.  Note how
the correlated insulator ($U=6$) has an exponential growth with thickness
as expected for a tunneling process, but it turns over at the
thickest junction, indicating a crossover to the incoherent transport regime.
The $U=5$ data, which is close to the critical point for a MIT, has neither
linear, nor exponential growth of its resistance-area product.  The
metallic cases ($U=2$ and 4) have perfect linear scaling of the resistance
with current, with a nonzero intercept corresponding to the contact
resistance. This may be surprising for $U=4$, because it is so strongly
scattering (with a mean free path much less than a lattice spacing), that
one would not think a semiclassical approach should apply there.
The constant satisfies $\sigma_0=2e^2/ha^2$.
}
\end{figure}

The Thouless energy is plotted versus temperature on a log-log plot for
$U=6$, which corresponds to a Mott-insulating barrier with a small
correlation-induced gap. The dashed line 
indicates where $E_{Th}=T$, which is an important crossover point for
dynamics, as we will see below.  Note that the temperature dependence is
significant in an insulator, because the integral in the denominator of
Eq.~(\ref{eq: thouless}) has strong temperature dependence in the insulator,
but the resistance does not in the tunneling regime at low temperature.
If we used the Thouless energy to determine the tunneling time via
$t_{tunnel}=\hbar/E_{Th}$, we would find tunneling times rapidly approaching
zero as $T\rightarrow 0$.  We will not comment further here as to whether there
is any substance to using such results to describe the quantum dynamics
of the tunneling process.  Instead we simply want to conclude that the
concept of the Thouless energy can be generalized to strongly correlated 
systems, and we will see below that the crossover point where $E_{Th}\approx T$
has important physical interpretations that will be developed in the next
section.  Finally, the generalization of the Thouless energy to
correlated systems changes the idea of a single energy scale being associated
with the transport, since now the energy scale develops strong temperature
dependence.  If a single number is desired, then we would propose to use
the energy scale where the Thouless energy is equal to the temperature,
indicated by the points of intersection of the solid lines with the
dashed curves in Fig.~\ref{fig: thouless_u=6}.

\section{Charge transport}

\begin{figure}[htbf]
\epsfxsize=3.0in
\epsffile{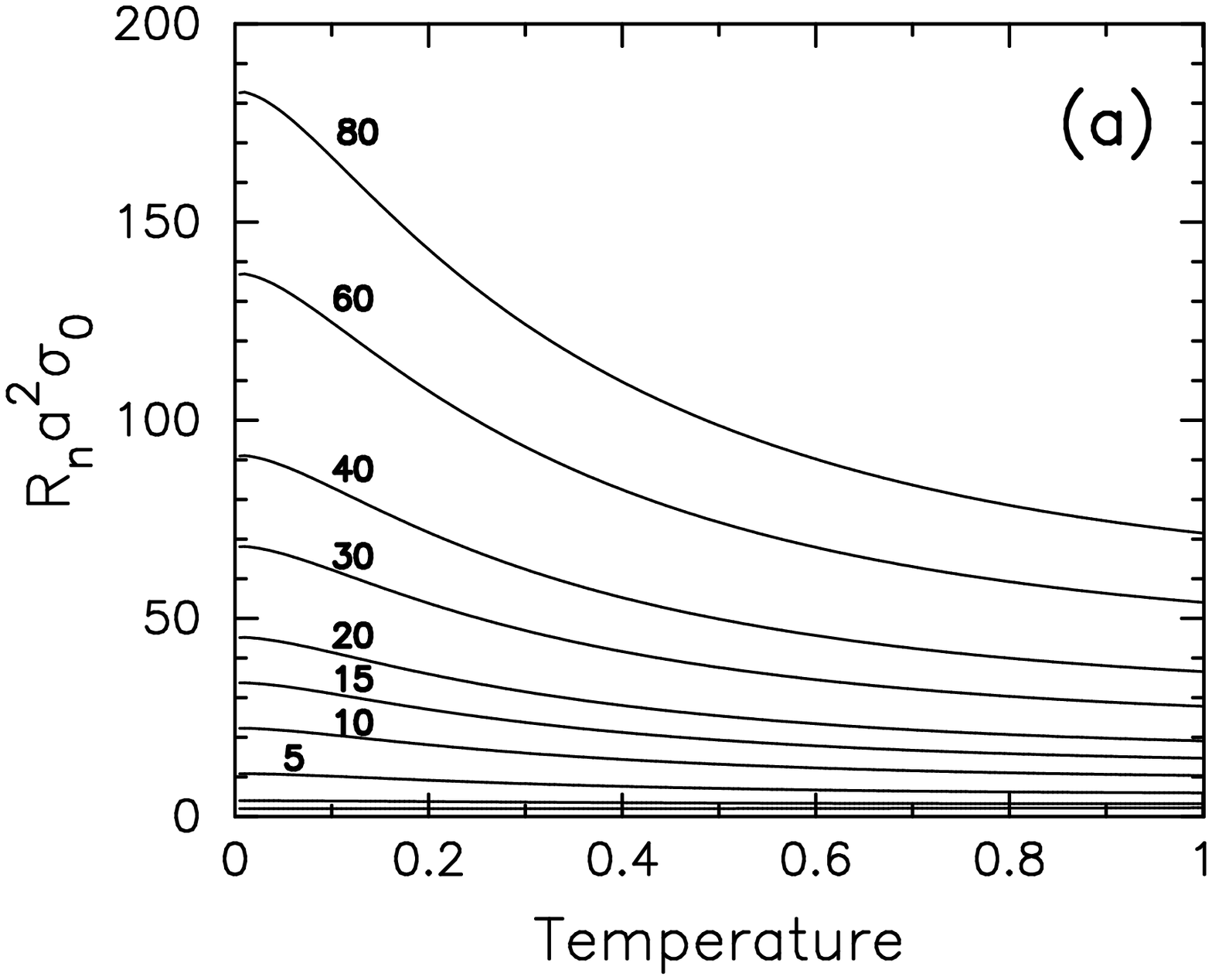}
\epsfxsize=3.0in
\epsffile{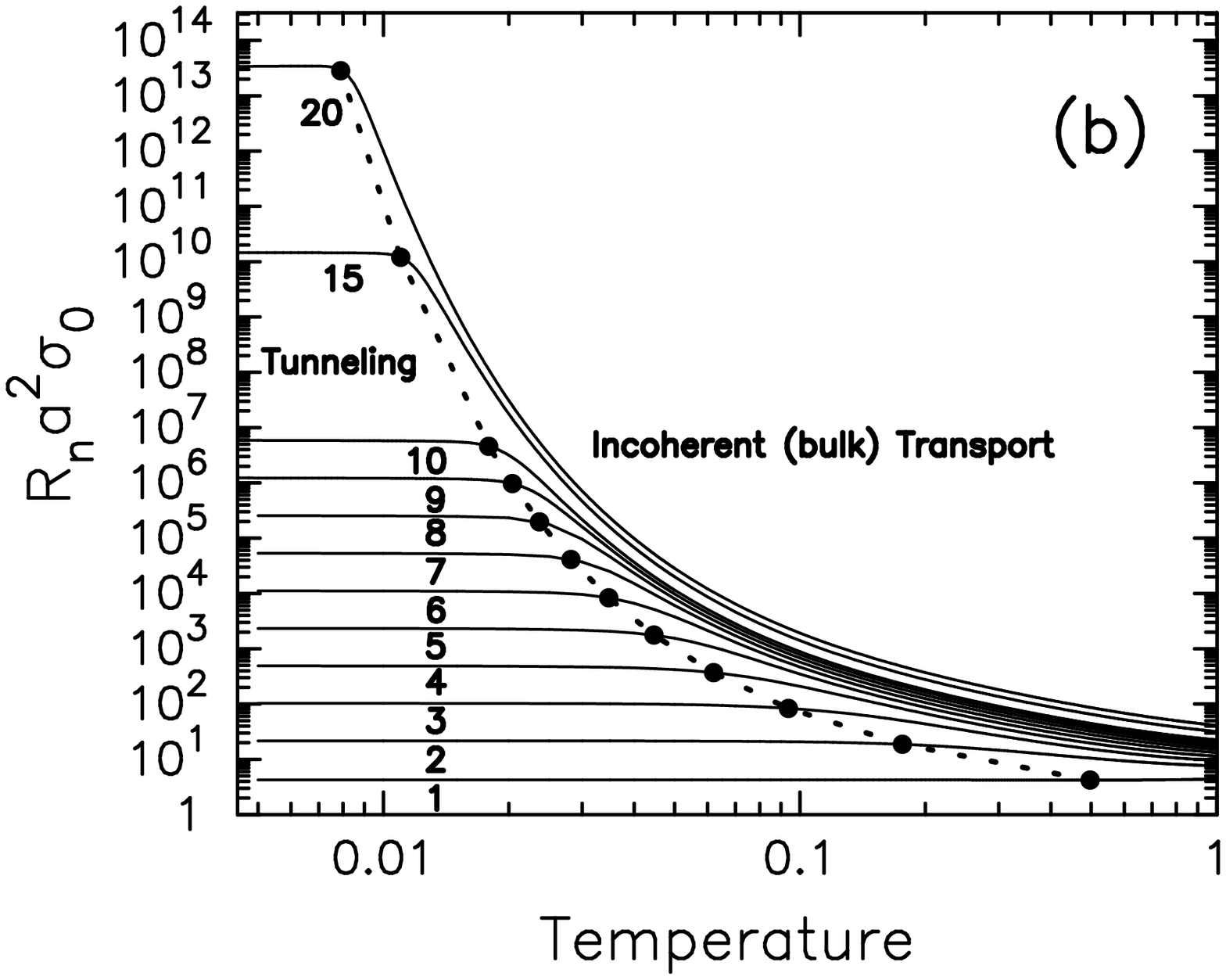}
\caption{\label{fig: rn_t}
Resistance-area product for nanostructures with (a) $U=4$ and (b) $U=6$
as a function of temperature [panel (a) is on a linear scale, and panel
(b) is a log-log plot].  In panel (a) we include results for
$N=1$, 2, (lowest two curves), 5, 10, 15, 20, 40, 60, and 80.  Note how
at each temperature there is a linear dependence of the resistance-area
product with the thickness of the junction.  Note further, that these junctions
have anomalous temperature dependence for a metal (they actually look
insulating in their dependence). In panel (b), we
show the results for $U=6$ with $N=1-10$, 15 and 20. Note at low
temperature we have tunneling, as the resistance-area product is
weakly dependent on temperature, and the steps are equally spaced as
a function of thickness,
indicating exponential dependence on the thickness. At higher temperatures,
there is a crossover to the incoherent transport regime, with the
resistance-area product picking up a strong $T$ dependence, and scaling
linearly with the thickness.  The dotted line that connects the solid
dots is a plot of the resistance-area value at the temperature
where $E_{Th}=T$ which determines the crossover.
}
\end{figure}

The dc resistance is a low-energy property of the nanostructure, and so
it requires the results of the single-particle properties to be determined
accurately at low energy.  This is not difficult for metallic barriers
with any degree of scattering, as long as the numerical subtleties
discussed above are taken into account in the analysis, but it does create 
problems for thick Mott insulators.  We need to be able to properly determine
the structure seen in Fig.~\ref{fig: self-energy} as the barrier is made
thicker, and this can exhaust the numerical resources, or the numerical
precision available for a given calculation.  For our work, we were not
successful in examining $U=6$ barriers thicker than $N=20$.

We plot the resistance-area product in Fig.~\ref{fig: rn_thick} for
$T=0.01$ and four different $U$ values: $U=2$, a diffusive metal near the 
Ioffe-Regel limit of a mean free path on the order of a lattice spacing;
$U=4$, a strongly scattering, anomalous metal, that has a strong dip in the
DOS near the chemical potential; $U=5$, a Mott-insulator that is nearly
critical; and $U=6$, a Mott-insulator with a small correlation-induced
gap. In panel (a), we have a semi-logarithmic plot, which is useful
for picking out tunneling behavior via an exponential increase of
the resistance with thickness.  This is clearly seen for the Mott
insulator with $U=6$, with the beginnings of a crossover occurring near
$N=20$, but the near-critical insulator at $U=5$ does not grow exponentially,
nor does it grow linearly [see panel (b)]. The data for $U=2$ and $U=4$,
both show linear increases with thickness, with a nonzero intercept on the
$y$-axis denoting the nonzero contact resistance with the metallic leads.
It is surprising that this linear ``Ohmic'' scaling holds for systems that
are so strongly scattering, that their mean free path is much less than one
lattice spacing.

Our final figure plots the resistance-area versus temperature for
(a) $U=4$ and (b) $U=6$ [Fig.~\ref{fig: rn_t}].  In panel (a), we
can infer a linear dependence of $R_na^2$ versus $L$ for all temperatures,
so this barrier is always Ohmic in nature.  But it has quite anomalous
temperature dependence, looking like an insulator, whose resistance is
reduced as the temperature increases. In panel (b), we see an exponential 
dependence of $R_na^2$ versus $L$ at low temperature, marked by the
equidistant step increases of $R_na^2$ as the thickness increases
(recall this is a log-log plot). The temperature dependence is also weak
in this region, indicated by the flatness of the curves.  Hence the system
is in the tunneling regime at low temperature.  As $T$ rises, there is a 
relatively sharp crossover region, where $R_na^2$ begins to pick up strong
(exponentially activated) $T$ dependence, and $R_na^2$ grows linearly
with $L$.  This is the incoherent ``Ohmic'' regime for the transport.
The solid dots represent the resistance-area product at the Thouless
energy, determined by finding the temperature where $E_{Th}=T$ from
Fig.~\ref{fig: thouless_u=6}, and marking those points on the curves
in panel (b).  A dashed line guide to the eye is drawn through these
points.  One can clearly see that the point where the Thouless energy
equals the temperature determines the crossover from tunneling to
incoherent transport.  Surprisingly, this crossover occurs at a
lower temperature for a thicker barrier.  This occurs, because the tunneling
resistance is higher for a thicker barrier.  As $T$ increases, the Ohmic 
resistance, determined by multiplying the temperature-dependent bulk 
resistivity by the thickness and dividing by the area, will decrease.
Once it is essentially equal to the tunneling resistance, there will
be a crossover from tunneling, which provides a 
``quantum short'' across the junction for low $T$, to ``Ohmic'' (incoherent)
thermally activated transport. This must occur at a lower temperature for 
more resistive junctions, and hence the thicker junctions have the
crossover before the thinner junctions.  Note that the temperature
scale for this crossover does not appear to have any simple relation to
the energy gap of the bulk material, instead it is intimately related
to the dynamical information encoded in the generalized $E_{Th}$
found in Eq.~(\ref{eq: thouless}).

We do not consider thermal transport there, since the thermopower vanishes 
for this particle-hole symmetric case and the thermal resistance is
not as interesting in systems with vanishing thermopower.

\section{Conclusions}

In this contribution we worked with a generalization of DMFT to inhomogeneous
systems to calculate the self-consistent many-body solutions for 
multilayered nanostructures that have barriers that can be tuned to
go through the Mott transition. We developed the computational formalism
thoroughly (based on the algorithm of Potthoff and Nolting), and although
we applied it only to the Falicov-Kimball model, it is obvious that one can
trivially add mean-field-like interactions such as Zeeman splitting for
magnetic systems, or long-range Coulomb interactions for systems with mismatched
chemical potentials.  In addition, one can invoke whatever impurity solver
desired for the local DMFT problem on each plane, which extracts a new self
energy from the current local Green's function.
We studied both the single-particle properties and the charge transport.

There are a number of interesting results that came out of this analysis.
First, we found that as the strength of the correlations increases in the
barrier, there is a stronger feedback effect on the Friedel-like oscillations
that appear in the metallic leads, but those oscillations vary little with the
thickness of the barrier for a fixed interaction strength. Second,
there are few oscillations inside the barrier except close to the 
interface with the metallic leads, but the behavior in the barrier,
of either an exponential decay, or of a constant DOS, gets frozen in
for a relatively thin barrier, and the DOS changes little with
increasing the thickness of the barrier, except when there is exponential decay
which will always decrease within the correlation-induced gap. Third,
the Mott insulating barrier develops a narrow peak-like structure in
the imaginary part of the self energy that approaches the bulk delta
function result.  This narrow and tall peak is difficult to determine
accurately with the numerics and limits the ability to study thick insulating
barriers. Fourth, we showed how to generalize the concept of a Thouless 
energy to become a function of $T$ for a strongly correlated Mott insulator.  
Our unifying form for the Thouless energy includes the results for both
the ballistic and diffusive metals as well.  We identified an
energy scale that describes the crossover from tunneling to incoherent transport
in these nanostructures; it corresponds to $E_{Th}=T$. This energy scale is 
quite useful in other areas such as in the theory of Josephson junctions,
which will be presented elsewhere. Sixth, we analyzed
the resistance of these devices and found interesting behavior, including
anomalous metallic behavior (but no tunneling) for a strongly scattering metal,
and the crossover from tunneling to Ohmic transport for insulating barriers.

This work also shed light on other approaches to transport through multilayered
structures like the Landauer-based approaches.  Usually these are 
non-self-consistent techniques that approach the problem from the point
of view of transmission and reflection of Bloch waves moving through
the device. We found that because the structure in the leads is frozen
in beginning with $N=1$ and because the exponential decay lengths are also
determined from $N=1$, if one knew those results and plugged them into
the Landauer approach, one should be able to calculate accurate properties;
i.e. the self consistency is needed for each nanostructure, but the 
self-consistency hardly changes with the thickness of the barrier. Hence
a phenomenological approach that adjusts the properties of the barrier
height to produce the required behavior, may work well, even for strongly
correlated systems; of course, the many-body theory is the only way
to determine the precise structure needed via its self-consistent solution
(i.e. it requires no fitting).

There are a number of important effects that we have not discussed here, which
play roles in the transport through nanostructures.  We did not attempt to 
include them in this first, simplest problem that we tackled.  The first
one is the issue of charge reorganization around the interface.  If the
chemical potentials of the leads and the barriers are different, electrons
will spill from one plane to the another until a screened dipole layer is 
formed, and a constant electrochemical potential is found throughout the 
device\cite{nikolic_freericks_miller_2002}.
Such effects can have dramatic results if one or more of the materials
is a correlated insulator, since the inhomogeneous doping of the system can
transform part of it from insulating to metallic.  This is believed to occur
in grain boundaries in high temperature superconducting tapes and 
wires\cite{hilgenkamp_mannhart_2002},
and in insulator-based 
nanostructures\cite{ohtomo_etal_2002,okamoto_millis_2004}.  
Second, calculations should be performed
off of half filling, where the thermal evolution of the chemical potential,
will likely undergo some temperature dependence so the charge rearrangement
can vary with temperature in the system.  Third, we should calculate the
thermal transport effects.  Since these calculations require particle-hole
asymmetry, we will have the chemical potential evolution and the charge
reorganizations to deal with as well.  Fourth, one can include ordered
phase effects at the mean-field level easily, as in a superconductor 
for a Josephson junction\cite{freericks_nikolic_miller_2002}, 
or in a ferromagnet for a spintronics device.
Fifth, it will be useful to determine the capacitance of a nanostructure,
since the capacitance is often important in determining the switching
speed of a device; it can be calculated with a linear-response formalism
as well.  Finally, we also should look into nonequilibrium effects, 
especially the nonlinear response of a current-voltage curve.  It is our
plan to investigate these complications in the future.

\section*{Acknowledgments}

We would like to thank V.~Zlati\'c for useful discussions.
We acknowledge support from the National
Science Foundation under grant number DMR-0210717 and the Office of Naval
Research under grant number N00014-99-1-0328. Supercomputer time was
provided by the Arctic Region Supercomputer Center and by the 
Mississippi Region Supercomputer Center ERDC.

\bibliography{fk_dmft.bib}

\end{document}